\documentclass[journal]{IEEEtran}
\usepackage{graphicx}
\usepackage{dcolumn}
\usepackage{bm}
\usepackage{stackrel}
\usepackage{amsthm}
\usepackage{amsmath}
\usepackage{amsfonts}
\usepackage{amssymb}
\usepackage{amssymb}
\usepackage{epstopdf}
\usepackage{longtable}
\usepackage{array}
\usepackage{latexsym}
\usepackage[latin1]{inputenc}
\usepackage{textcomp}
\usepackage{subfigure}
\usepackage{braket}
\usepackage{wasysym}
\usepackage{ulem}
\usepackage{color, soul}
\usepackage{pifont}
\usepackage{multirow}
\usepackage[noadjust]{cite}

\ifCLASSINFOpdf

\else
\fi

\begin{document}

\title{Non-linear frequency-sweep correction of tunable electromagnetic sources}
\author{\IEEEauthorblockN{M. Minissale\IEEEauthorrefmark{1}\IEEEauthorrefmark{2}\IEEEauthorrefmark{3},
T. Zanon-Willette\IEEEauthorrefmark{1},
I. Prokhorov\IEEEauthorrefmark{4},
 H. Elandaloussi\IEEEauthorrefmark{1}, and
C. Janssen\IEEEauthorrefmark{1}\\
\IEEEauthorblockA{\IEEEauthorrefmark{1}Sorbonne Universit\'e, Observatoire de Paris, Universit\'e PSL, CNRS, LERMA, F-75005, Paris, France }\\
\IEEEauthorblockA{\IEEEauthorrefmark{2}Aix Marseille University, CNRS, PIIM, Marseille, France}\\
\IEEEauthorblockA{\IEEEauthorrefmark{3}Aix Marseille University, CNRS, Centrale Marseille, Institut Fresnel, Marseille, France }\\
\IEEEauthorblockA{\IEEEauthorrefmark{4}Institute of Environmental Physics, Heidelberg University, Im Neuenheimer Feld 229, 69120 Heidelberg, Germany}
\thanks{Manuscript received February xx, 2018; revised xxx 2018.
Corresponding author: M. Minissale (email: marco.minissale@univ-amu.fr).}}}

\markboth{IEEE Transactions on Ultrasonics, Ferroelectrics, and Frequency control}%
{Shell \MakeLowercase{\textit{et al.}}: Bare Demo of IEEEtran.cls for IEEE Journals}

\maketitle

\begin{abstract}
Tunable electromagnetic sources, such as voltage controlled oscillators (VCO), micro electromechanical systems (MEMS) or diode lasers are often required to be linear during frequency-sweep modulation. In many cases, it might also be sufficient that the degree of the non-linearity can be well controlled. Without further efforts, these conditions are rarely achieved using free running sources.
Based on a pre-distortion voltage ramp, we develop in this letter a simple and universal method that minimizes the non-linear frequency response of tunable electromagnetic sources. Using a current-driven Quantum Cascade Laser (QCL) as an example, we demonstrate that the non-linearity can easily be reduced by a factor of ten when using a single distortion parameter $\gamma$. In the investigation of the IR absorption spectrum of ozone at 10\,$\mu$m, an even better reduction of the frequency scale error by two orders of magnitude is obtained by using the pre-distortion method to generate an essentially purely quadratic sweep frequency dependency which can be inverted easily to retrieve precise molecular line positions. After having tested our method on a variety of electromagnetic sources, we anticipate a wide range of applications in a variety of fields.
\end{abstract}

\begin{IEEEkeywords}
Tunable lasers, Frequency sweep,  Non-linearity, Gamma correction, QCL.
\end{IEEEkeywords}

\IEEEpeerreviewmaketitle

\section{Introduction}
\label{intro}

Tuneable electromagnetic (EM) sources are now available covering all the frequency ranges from RF to THz. Specialized applications in various fields, such as radar detection \cite{Wang2011}, pollutant monitoring \cite{Kleinert06}, infrared countermeasures (IRCM) in military defence \cite{Bradshaw2009}, laser surgery \cite{AtezhevBarchunovVartapetovEtAl2016} and medical diagnostics or tomography \cite{JovicichCzannerGreveEtAl2006,Klein:2017ik}, require that these light sources have linear tuning rates.

In the RF and HF domains for example, voltage controlled oscillators (VCO) are widespread and offer good tunability. They are thus intensively used in radar applications with frequency modulated continuous wave (FMCW) measurements ranging from target intrusion \cite{Butler2007}, over medical monitoring \cite{van_loon_wireless_2016} to snow depth studies on Antarctic sea ice [9]. However, VCOs suffer from non-linearities which deteriorate the spatial resolution. These problems are well documented and are usually solved by various types of compensation techniques based on hardware corrections with complex pre-distorted voltage ramps, phase-locked loop devices or software analysis \cite{Iiyama1996,Ahn2007,Jung2013,Lazam2015}.
Nevertheless those methods are often too complex for an easy implementation and new simple techniques have to be considered.

In the IR spectral ranges, optical parametric oscillators (OPOs) are nowadays becoming  efficient tunable mid-IR laser sources. Recent progress \cite{Courtois2013, Andrieux2011} has led to sweeps of up to 75\,cm$^{-1}$ ($\sim$ 2 THz) at THz frequencies with a singly resonant OPO. Such a sweeping range leads to a large non-linear frequency response.
Quantum-cascade (QC) and inter-band cavity (IC) lasers have also proven to be powerful sources for mid-IR wavelengths and can also provide large tuning ranges \cite{HugiMauliniFaist2010,Chen:2014dp,Bidaux:2015jl}. They are based upon multiple quantum well semiconductor lasers with high power outputs at room temperature and the development of external cavity QCLs will offer a strong improvement in high-resolution spectroscopy of large complex molecules in the gas or liquid phases \cite{Courtois2013,Lambrecht2014}.
Once again, theses sources with tuning ranges of a few cm$^{-1}$ also suffer from non-linearities and in molecular spectroscopy one often employs a post-analysis based on local frequency markers, such as Fabry-P\'erot-etalons (FPE).

\indent We present, in this paper, a simple technique for the reduction of laser frequency sweep errors, taking accurate molecular spectroscopy as an example. We have tested and validated the linearization technique for different laser sources:
\begin{itemize}
\item a distributed feedback (DFB) Quantum Cascade Laser (QCL) from \textit{Alpes Laser} emitting at 9.54 $\mu$m;
\item a DFB-Interband Cascade Laser (ICL) from \textit{Nanoplus} emitting at 4.44 $\mu$m;
\item a high voltage Lead zirconate titanate (PZT) ceramic coupled to an external cavity diode laser (ECDL) from \textit{New Focus} emitting at 1.06 $\mu$m.
\end{itemize}
Our linearization approach is inspired by the well-known $\gamma$-correction technique applied in video and TV technology \cite{Hoang2010}. We propose realizing a non-linear voltage ramp as a pre-distortion signal, which  drives the laser source for controlling the degree of non-linearity. The pre-distortion ramp, that contains only a single control parameter $\gamma$, can be applied in two different ways: to directly reduce the non-linear frequency response of the laser source by one order of magnitude (\textit{method 1}); to make the frequency response highly quadratic.
Combined with a subsequent and rapid (analytical) inversion of the quadratic polynomial, the non-linearity is reduced by a factor of 100 or more over the entire frequency sweep range (\textit{method 2}).

\section{Results}

\subsection{Method 1: direct non-linear reduction}
\begin{figure}[b!!]
\centering
\includegraphics[width=8.7cm]{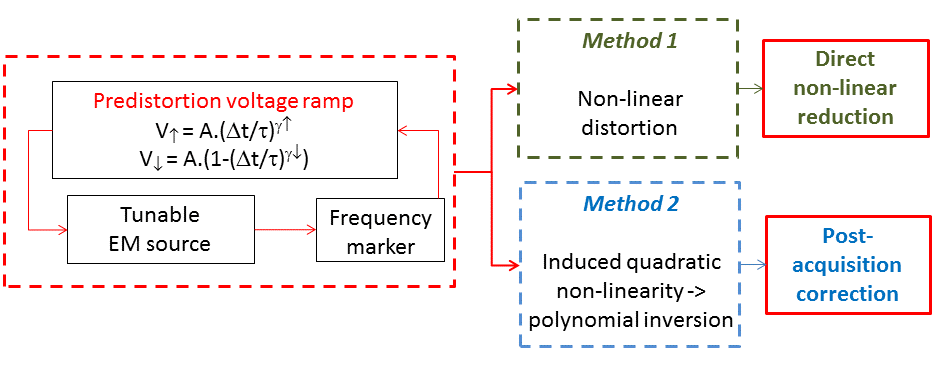}
\caption{(Color online) Block diagrams for the non-linear frequency sweep correction. Possible application methods: either direct reduction of the sweep non-linearity (method 1) or post acquisition treatment by polynomial inversion for non-linear frequency sweep correction (method 2).}
\label{fig:NL-scheme}
\end{figure}
Fig.~\ref{fig:NL-scheme} shows the schematic block diagram of the experimental procedure for controlling the non-linearity of the electromagnetic (laser) source. The results presented in this paper are obtained using a DFB-QCL, even if similar results have been obtained for other laser sources. The DFB-QCL is centered at 1049~cm$^{-1}$ (9.54\,$\mu$m) and emits in the wavenumber range from 1046.2 to 1053.1~cm$^{-1}$. The QCL can be operated at temperatures between $-25$ and 20$\,^{\circ}$C and tolerates a maximum current of 1.48\,A. Its  threshold current is  0.88\,A at $-25\,^{\circ}$C and 1.31\,A at 20$^{\circ}$C. Under our operation conditions the QCL emits a  power of $\sim$ 25\,mW and has a tuning rate of $\partial\nu/\partial I\sim 5.6 \cdot 10^{-3}$\,cm$^{-1}$/mA (170\,MHz/mA). The current was provided by a commercial source (\textit{LDX-3232}, \textit{ILX-Lightwave}).

The emitted laser beam is collimated and size-reduced by a germanium telescope and then split by a 50-50\% beamsplitter (BS). The splitted light beam either traverses an absorption cell filled with ozone (O$_{3}$) or passes a FPE. The absorption cell is a Teflon coated steal cell equipped with two wedged BaF$_2$ windows. The cell has a base length of roughly 40 cm and a diameter of 50 mm. The two output beams from the cell and from the FPE are focused onto two HgCdTe detectors. The photo-currents of these two detectors are first pre-amplified and then recorded by a multi-purpose data acquisition card (DAQ, \textit{NI PCI-6281}) onboard a PC. The DAQ card is also used to generate an adjustable voltage ramp for modulating the laser current and sweeping the QCL emission frequency.
The modulation signal is one out of the following:
\begin{subequations}
\begin{align}
V_{\uparrow}= &\ A \left(\frac{\Delta t}{\tau}\right)^{\gamma_{\uparrow}}\label{eqV1} \quad &\text{(ramp up)}\\
V_{\downarrow}= &\ A\left[1-\left(\frac{\Delta t}{\tau}\right)^{\gamma_{\downarrow}}\right] \quad & \text{(ramp down)}\label{eqV2}\\
V_{\uparrow\downarrow}= &\  A\ \left\{  \left(\frac{2 \Delta t}{\tau}\right)^{\gamma_{\uparrow}}
\Theta \left(\tau - 2 \Delta t\right)+\right.  \nonumber \\
& \left. \left[1-\left(\frac{2 \Delta t -\tau}{\tau} \right)^{\gamma_{\downarrow}}\right] \Theta \left(2 \Delta t -\tau \right) \right\} \label{eqV3}  &\text{(triangular)}
\end{align}
\end{subequations}
where $\tau$ is the ramp duration, $\Delta t$ one incremental time step in the ramp of $N=\tau/\Delta t$ steps, $A$ the amplitude and $\gamma_{\uparrow}$, $\gamma_{\downarrow}$ two arbitrary exponents $>0$. $\Theta (x)$ is the Heavyside step function, which takes $ 1$ for $x> 0$, 1/2 for $x= 0$ and 0 otherwise.
\begin{figure}[b!!]
\centering
\includegraphics[width=9cm]{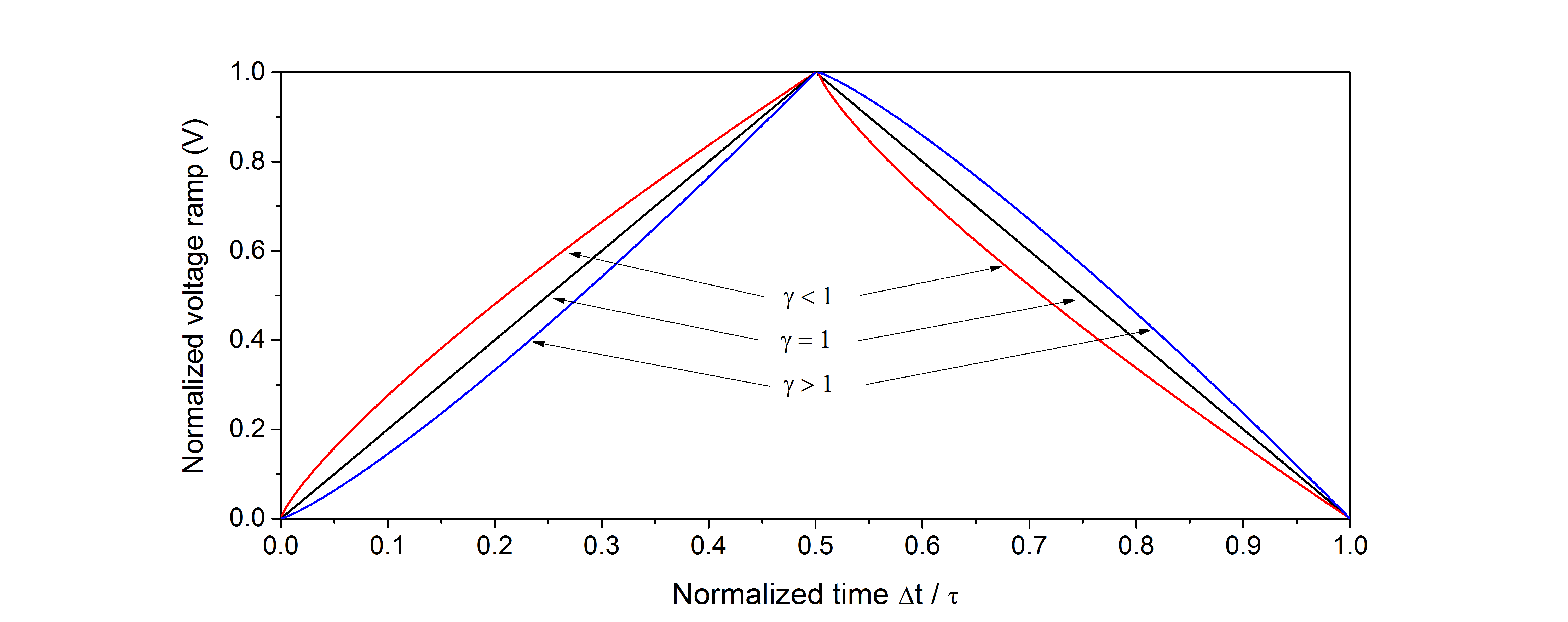}
\caption{(Color online) Examples of pre-distortion ramps applied to the tunable electromagnetic source. Left: ramp up; right: ramp down.}
\label{fig:ramp}
\end{figure}
In the case of our current source, the voltage to current transfer function was given by $dI/dV=200\,$mA/V. For the sake of clarity, we present only results obtained with the positive ramp $V_{\uparrow}$, even if similar non-linearity corrections have been obtained with the down voltage ramp $V_{\downarrow}$. Without loss of unambiguity, we can therefore drop the $\uparrow, \downarrow$ index notation in what follows.
We show in Fig.~\ref{fig:ramp} three possible curves for the pre-distortion voltage ramp $V$: $\gamma$-correction factors $\gamma=1$, $\gamma<1$ and $\gamma>1$. These voltage ramps are applied to the current driver and allow tuning for approximately 230\,mA or 1.7\,cm$^{-1}$ without any mode hopping.

\begin{figure*}[t]
\centering
\includegraphics[width=15cm]{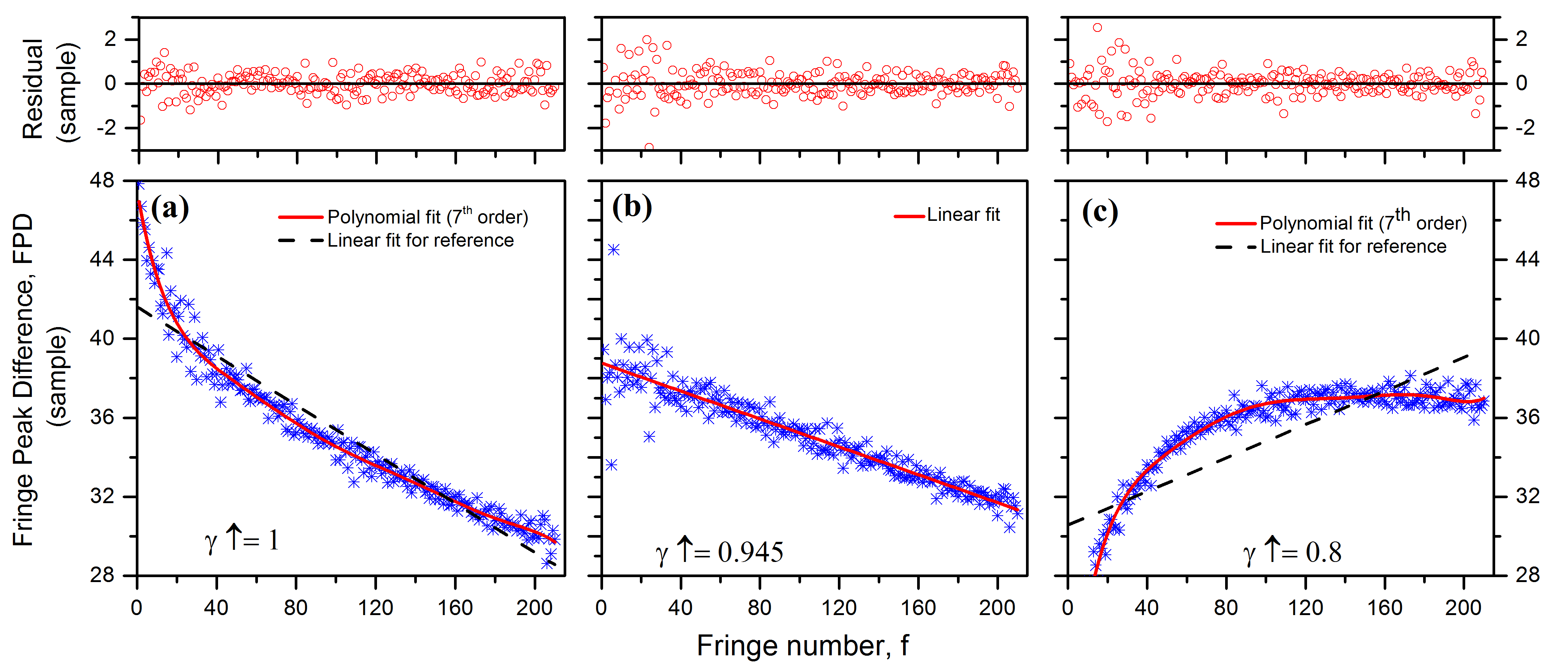}
\caption{(Color online) Fringe peak difference (FPD) versus fringe number for various $\gamma$-correction factors: (a) $\gamma=1$; (b) $\gamma=0.945$; (c) $\gamma=0.8$. The polynomial order is either 7 (a, c) or 1 (b). The residual sample error obtained from the polynomial fit is shown in the uppermost traces above the graphs.}\label{fig:Nl-ramp}
\end{figure*}
We demonstrate first that a linear ($\gamma = 1$) voltage ramp leads to a non-linear frequency scale and that a modification of the voltage ramp with a tunable distortion based on Eq.~\eqref{eqV1} (or \eqref{eqV2}) is able to efficiently compensate this discrepancy or to generate a non-linear quadratic dependence on the laser frequency scan.
To study the impact of the distortion parameter, we use a frequency discriminator based on a low finesse FPE (\textit{AL6000, AeroLaser}, $F\sim 5.5$), which has a fringe width of $ 1.5 \cdot 10 ^{-3}$\,cm$^{-1}$ (45\,MHz) and a free spectral range (FSR) of $\delta\nu =  8.0 \cdot 10^{-3}$\,cm$^{-1}$ (240\,MHz).
A typical Fabry-P\'erot spectrum thus contains 150 to 200 fringes, from which the frequency range can be obtained by counting the number of fringes in our spectra. We also determine the fringe peak difference (hereafter FPD) between two consecutive fringes expressed in spectral point number $i$. The results are reported in Fig.~\ref{fig:Nl-ramp}(a) to (c), where three different ramp forms $\gamma=1$ (a), $\gamma=0.945$ (b), and $\gamma=0.8$ (c) are compared by means of the FPD signal. In order to characterize their frequency dependence, the FPD signals are fitted by polynomial functions of up to the seventh order when necessary, with the corresponding residuals plotted on top of the FDP curves.

Clearly, we observe that the FPD signal is strongly non-linear as a function of the frequency scale (fringe number), depending on the selected $\gamma$-correction factor, and for $\gamma = 1$ in particular. It is interesting to observe that the FPD curve changes from a concave to a convex shape when going from high to low values of $\gamma$. It is particularly remarkable that at an intermediate value $\gamma=0.945$, a perfectly linear dependency is achieved. The residuals of the linear fit are as low as those obtained from fitting polynomials of a much higher degree to curves reported in Fig.~\ref{fig:Nl-ramp}(a) and Fig.~\ref{fig:Nl-ramp}(c).
A constant FPD (with zero slope) would be very convenient, because it signifies a perfectly linearized spectrum. However, our tests using many different scenarios with various $\gamma$ correction values or other ramp forms showed that a zero slope deviation could never be realized.

\begin{figure}[t]
\centering
\includegraphics[width=9cm]{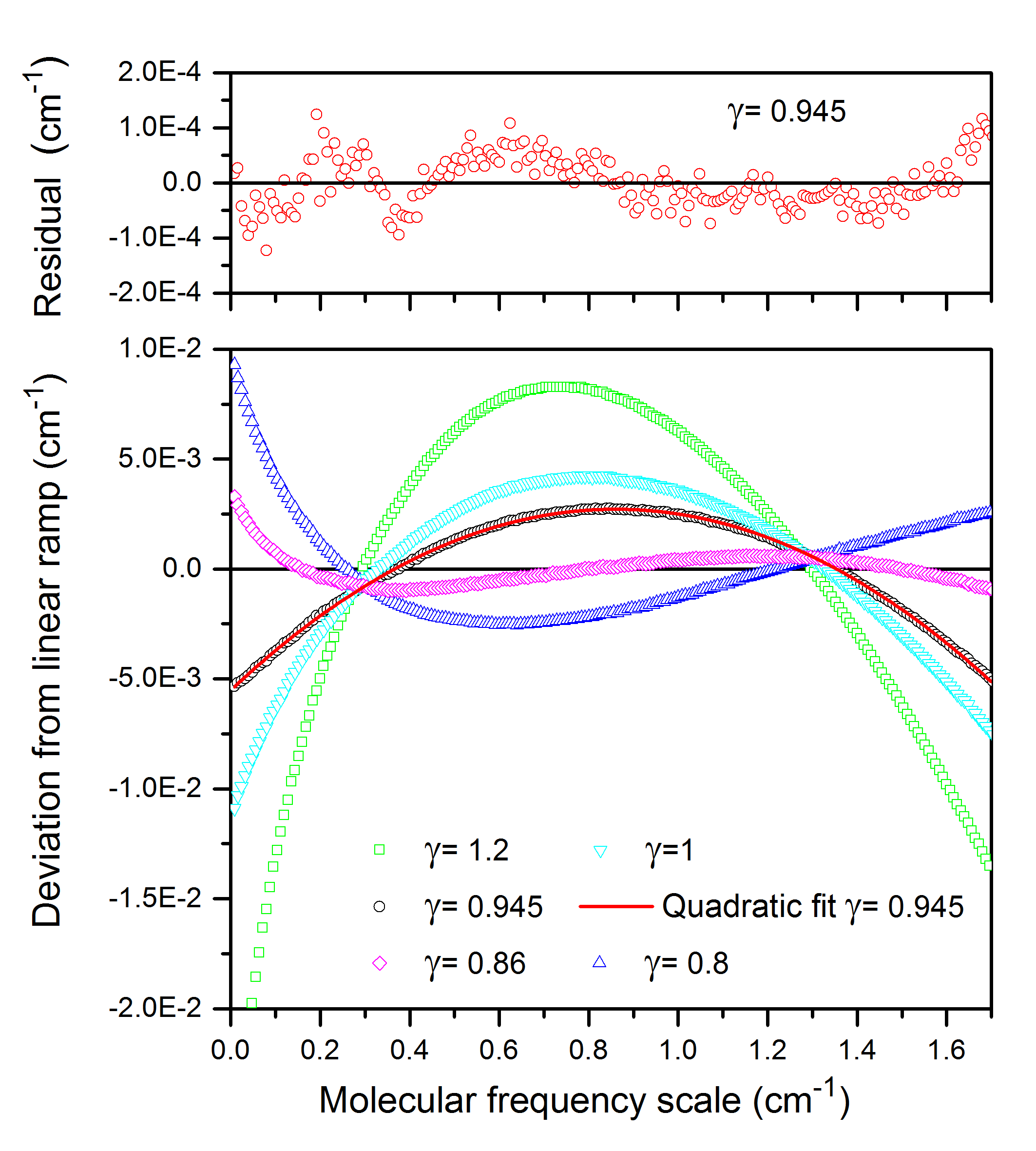}
\caption{(Color online). Wave number error between a linear frequency sweep and the non-linear sweep generated with $\gamma=1.2$, 1, 0.945, 0.86, and 0.8. Top panel shows the residual wave number  over the entire frequency scale using the simple quadratic fit of results with $\gamma=0.945$ (black circles).}
\label{fig:quadratic-ramp}
\end{figure}

Nevertheless, we have been able to systematically reproduce a parabolic frequency response which yielded a linearly decreasing FDP ($\gamma=0.945$ in Fig.~\ref{fig:Nl-ramp}(b)). In this case ${d\nu}/{df} = const.$, where $f$ is the fringe number.
To illustrate this point further, we plot the deviation of etalon peak positions from a linear rate for five different values of $\gamma=$1.2, 1, 0.945, 0.86, and 0.8 in Fig.~\ref{fig:quadratic-ramp}. As expected, we find a quasi-parabolic behavior in the case of $\gamma=0.945$.
We note that for a value of $\gamma=0.86$, we are able to reduce the non-linear frequency response of the laser source (\textit{method 1}) without any post-processing of data. Indeed, this $\gamma$ value is not able to completely eliminate the non-linearity, but reduces the non-linear frequency response to values lower than $3 \cdot 10^{-3}$\,cm$^{-1}$ (90 MHz) over the whole tuning range or to lower than $1.5 \cdot 10^{-3}$\,cm$^{-1}$ (45 MHz) if one considers only 90\% of the sweep range.

\subsection{Method 2: Post-acquisition correction}

In order to control the non-linear frequency response even better, we now propose an efficient frequency correction and post-processing scheme (\textit{method 2}): first, a quadratic frequency response (i.e. $\gamma=0.945$ for the ramp-up of the DFB-QCL) in the spectral acquisition is generated. The complete linearization of the frequency scale is then obtained by analytic solution of the quadratic equation describing the frequency dependence. In table~\ref{tab:2} we list the full set of gamma correction factors used in Eqs.\ref{eqV1} and \ref{eqV2} to have a quasi-parabolic frequency response for the three laser sources (QCL, ICL, and PZT-ECDL) both for the up and the down voltage ramp.
\begin{table}[th]
\centering
\caption{Gamma correction factors for different laser sources.\label{tab:2}}
\footnotesize
\begin{tabular}{c c c c c}
\hline
\hline
Laser & Wavelength	& Tuning range &  $\gamma_{\uparrow}$ & $\gamma_{\downarrow}$ \\
source & $\mu$m	     &  cm$^{-1}$ & 	           &        \\
\hline
DFB-QCL$^a$	&	9.54	&	2 & $0.945$  & $0.958$	\\
DFB-ICL$^b$	&	4.44	& 2 & $0.450$  & $0.500$		\\
PZT-ECDL$^c$	&			1.06	& 1 & $0.450$  & $0.375$\\
\hline
\hline
\multicolumn{5}{l}{$^a$Distributed feedback Quantum Cascade Laser.} \\
\multicolumn{5}{l}{$^b$Distributed feedback Interband Cascade Laser .}\\
\multicolumn{5}{l}{$^c$Lead zirconate titanate ceramic coupled to an external cavity diode laser.}

\end{tabular}
\end{table}

We have already shown that a quadratic frequency response corresponds to a linear decrease  of the FPD signal. The offset $b$ and slope $m$ values of the FPD line can be determined experimentally from Fig.~\ref{fig:Nl-ramp}(b). These two parameters allow to easily convert from the time to the frequency domain and provide a very accurate frequency scale. In the following, we will discuss in detail how this is done and how this reduces non-linearity induced frequency errors by a factor of 100 over the entire frequency sweep.
When using a FPE with a free spectral range $\delta\nu$, we can record the FPD signal versus the fringe number \textit{f}, which defines the following frequency scale:
\begin{align}\label{eq0}
 \nu=\nu_0+f\cdot\delta\nu.
\end{align}
If we measure the sample number difference $\Delta i$ between two acquired fringe extrema (FPD), we find that $\Delta i/\Delta f$ can be linearized in \textit{f} as follows
\begin{align}\label{eq:de}
 \frac{\Delta i}{\Delta\nu}=\frac{1}{\delta\nu}\frac{\Delta i}{\Delta f}=\frac{1}{\delta\nu}(b+m\cdot f),
\end{align}
using our particular value of $\gamma$. Here, $i$ is again the point number and $\Delta i/{\Delta\nu}$ is the (inverse) frequency tuning rate. Since $\Delta i/\Delta f$ is a linear function in $f$, it can be identified with the derivative $\Delta i/\Delta f  = d i/d f$. This makes Eq.~\eqref{eq:de}
a differential equation that can be integrated to yield the point number $i$ as a quadratic function in $\nu$.

This relation can then easily be inverted to give the frequency as a function of spectral point number:
\begin{align}\label{eq:linscale}
 \nu(i)-\nu_0=\frac{\delta\nu}{m}(\sqrt{b^2+2m\cdot i}-b),
\end{align}
where the integration constant has been eliminated by fixing the frequency of the initial point. We note that it is not necessary to know the free spectral range ${\delta\nu}$ of the FPE beforehand. It is only required that markers of the frequency scale are provided, which allow to determine the constants $b$ and $m$ in Fig.~\ref{fig:Nl-ramp}(b). If two frequencies in the spectrum are known, the scale factor ${\delta\nu}$ can be easily obtained a-posteriori after the linearization. It is also not required the frequency discriminator to be a marker with regular frequency spacing, such as an FPE. One could also use a sufficient number of well known absorption lines that have an irregular frequency spacing, even though this evidently is slightly less convenient. In this case, one would directly work on the frequency scale $\nu$ (instead of $f$) and set $\delta \nu = 1 $ in Eq.~\eqref{eq:linscale}. For the method to work, it is only required that the polynomial shape of the $\Delta i/\Delta \nu$ curve can be retrieved, that the linear case can be identified and that the parameters $b$ and $m$ can well be determined. Note that we have expressed our calculation in terms of the point number $i$, but since points are acquired at constant time intervals $t=i\Delta t$, the corresponding time domain equation only contains an additional scaling factor of $\Delta t^{-1}$.
In practice, one might well live with the quadratic frequency dependence of Eq.~\eqref{eq:linscale}, as it already provides a highly accurate frequency scaling. However, one can also numerically resample the spectrum based on the accurate quadratic frequency dependence, which leads to a truly linearized solution at the additional cost of an interpolation.

\subsection{Spectroscopic application}

\begin{figure}[t]
\centering
\resizebox{9cm}{!}{\includegraphics[angle=0]{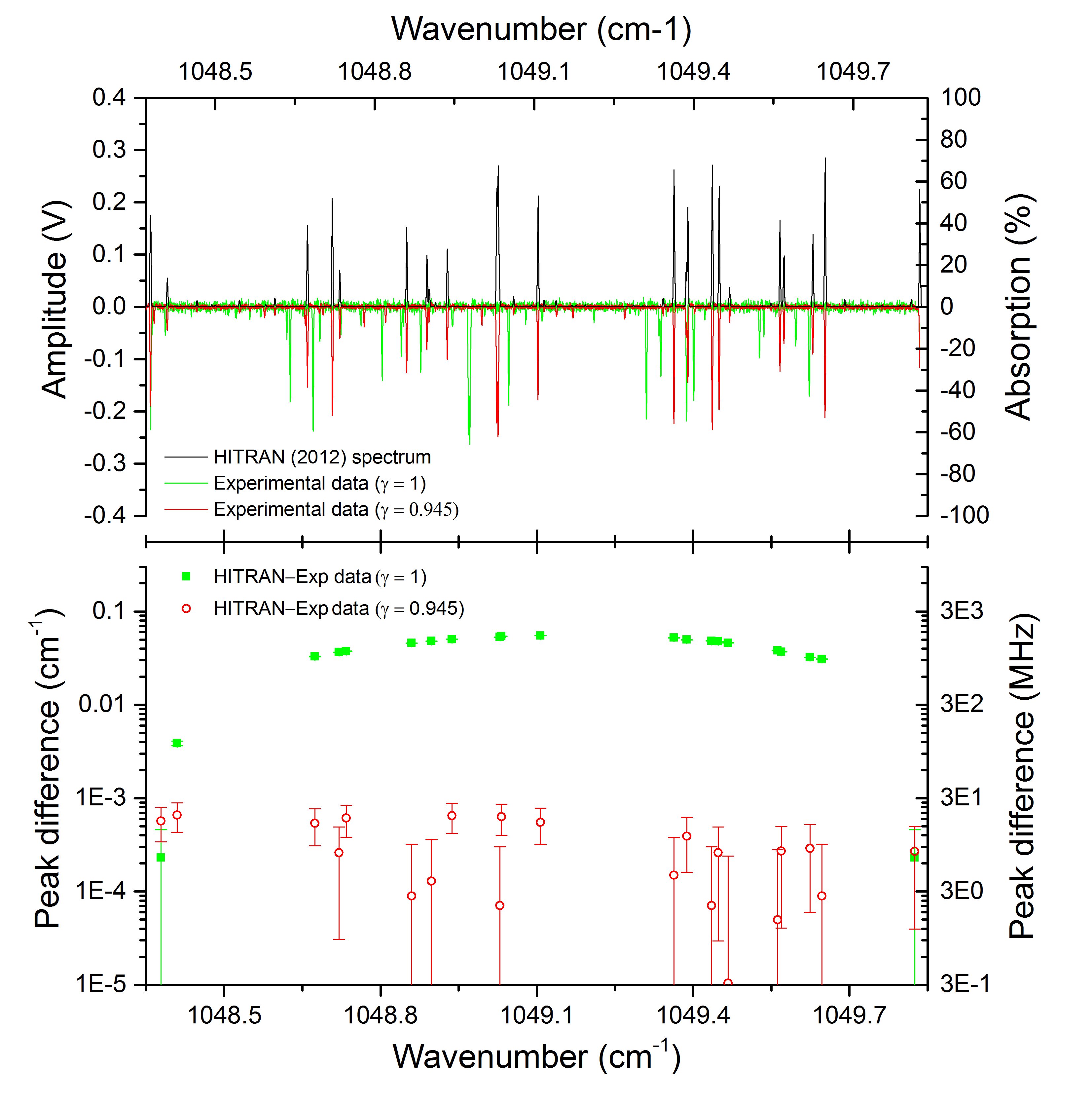}}
\caption{(Color online) Top panel: synthetic (HITRAN 2012, black curve) and experimental $^{16}$O$_{3}$ spectra recorded with a gamma factor $\gamma=$ 1 (green curve) and 0.945 (red curve). Spectrum recorded with $\gamma= 0.945$  is shown after application of \textit{method 2}. Bottom panel: line position difference between predicted and recorded ($\gamma= 1$  and 0.945 respectively green and red dots) line frequency.}
\label{fig:corrected-spectrum}
\end{figure}

We have thus shown how to linearize the frequency scale by imposing a well-known non-linearity on the voltage ramp. We will now present a direct spectroscopic application. In particular, we apply our method to the acquisition of IR absorption spectra of ozone. These spectra are obtained at 25$^{\circ}$C by filling the absorption cell with about 100~mTorr of O$_3$. The temperature of the QLC is set to $-10\,^{\circ}$C and the base current to 1.1~A. We sweep the laser frequency by applying a voltage ramp with $A \sim 1.45\,$V, corresponding to a current sweep of about 290~mA. We are able to identify the spectral range by comparison with a synthetic spectrum from the HITRAN database, shown by the black spectrum in the upper top panel of Fig.~\ref{fig:corrected-spectrum}.
The maximal discrepancy between predicted HITRAN \cite{RothmanGordonBabikovEtAl2013} and experimental line frequency positions (obtained with the unmodified linear voltage ramp) is between 4 to $6\cdot 10^{-2}$~cm$^{-1}$ ($\sim1.2-1.8$~GHz) over 1.4~cm$^{-1}$. This is indicated by green dots in the bottom panel of Fig.~\ref{fig:corrected-spectrum}.
The red dots in the bottom panel of Fig.~\ref{fig:corrected-spectrum} show that the implementation of \textit{method 2} (with $\gamma=0.945$) is able to reduce the line position discrepancy between predicted and recorded line frequency positions to less than $6 \cdot 10^{-4}$~cm$^{-1}$ ($\sim 12$~MHz) over the same frequency sweep range, thus effectively reducing the non-linearity level by a factor of 100 or better.

\section{Conclusion}
In the present paper, we have presented a simple and universal method to reduce or correct the non-linear frequency sweep behavior of free-running lasers. Most accurate frequency control is achieved when the pre-distortion is used to force a quadratic frequency dependence that can easily be accounted for by straight-forward post-data processing.
We expect our technique for non-linear frequency sweep correction to have a wide range of applications in  many other fields, such as atomic and molecular high resolution spectroscopy, radar sensing, MEMS and medical diagnostics.\\
Presently we do not attempt to explain the physical phenomena behind the non-linearity parameter $\gamma$, but we note that our simple one parameter ansatz in Eq. (1) has a constant logarithmic derivative, implying that fractional changes of the time variable are proportional to fractional changes of the amplitude. We have tested this method for different sources working in different frequency domains, as a DFB-ICL at 4.44 $\mu$m and a PZT-ECDL at 1.06 $\mu$m and we have always found similar results when applying our \textit{method 2}. We note that $\gamma$-correction values are not universal, but strongly depend on the laser, the current source, the (electro-optical or opto-mechanical) driving mechanism, etc.  Nevertheless, these parameters can be easily found by using a suitable frequency discriminator (e.g. Fabry-P\'erot etalon, molecular line markers, etc.).

\section*{Acknowledgment}

\indent This work was supported by grants from R\'egion Ile-de-France in the framework of the DIM ACAV and by the LABEX Cluster of Excellence FIRST-TF (ANR-10-LABX-48-01), within the Program Investissements d'Avenir operated by the French National Research Agency (ANR).

\end{document}